# Protein accumulation in the endoplasmic reticulum as a non-equilibrium phase transition

Zoe Budrikis[1], Giulio Costantini[2], Caterina A.M. La Porta[3] & Stefano Zapperi[1,2]

Several neurological disorders are associated with the aggregation of aberrant proteins, often localized in intracellular organelles such as the endoplasmic reticulum. Here we study protein aggregation kinetics by mean-field reactions and three dimensional Monte carlo simulations of diffusion-limited aggregation of linear polymers in a confined space, representing the endoplasmic reticulum. By tuning the rates of protein production and degradation, we show that the system undergoes a non-equilibrium phase transition from a physiological phase with little or no polymer accumulation to a pathological phase characterized by persistent polymerization. A combination of external factors accumulating during the lifetime of a patient can thus slightly modify the phase transition control parameters, tipping the balance from a long symptomless lag phase to an accelerated pathological development. The model can be successfully used to interpret experimental data on amyloid-β clearance from the central nervous system.

[1] Institute for Scientific Interchange Foundation, Via Alassio 11/C, Torino 10126, Italy. [2] Istituto per l'Energetica e le Interfasi, CNR-Consiglio Nazionale delle Ricerche, Via R. Cozzi 53, Milano 20125, Italy. [3] Department of Biosciences, University of Milano, via Celoria 26, Milano 20133, Italy. Correspondence and requests for materials should be addressed to S.Z. (email: stefano.zapperi@cnr.it).

ARTICLENATURE COMMUNICATIONS | DOI: 10.1038/ncomms4620

A number of neurological pathologies, including Alzheimer's disease (AD) and Parkinson's disease, spongiform encephalopathies and serpinopathies, are collectively identified as conformational diseases because they are all characterized by the aggregation and tissue deposition of aberrant conformations of proteins. In recent years, a large effort has been devoted to understanding the biological mechanisms underlying the biochemical properties of these proteins and the associated secretory pathways. While the key mechanisms triggering these diseases are still largely unknown, recent evidence points towards a pivotal role played by secretory pathways that is common to all these pathologies.

Serpinopathies result from point mutations in α1-antitrypsin and neuroserpin showing a delay in folding, with unstable intermediates being cleared by endoplasmic reticulum (ER)-associated degradation (ERAD)[1–3]. The remaining proteins are either fully folded and secreted or retained as ordered polymers within the ER of the cell of synthesis. Mutations in neuroserpin result in the autosomal dominant inclusion body dementia (Familial Encephalopathy with Neuroserpin Inclusion Bodies (FENIB))[4], and mutants of neuroserpin in FENIB patients show accelerated rates of polymerization compared with wild type protein both at the protein level[5,6] and in cell models[7,8]. The current picture of these diseases shows that accumulation of misfolded proteins within the lumen of the ER activates ERAD pathway. When ERAD fails to remove sufficient mutants, the remaining proteins start to polymerize[7,9,10].

In spongiform encephalopathies, recent evidence shows a new pathway that contributes to prion propagation owing to proteasomal dysfunction and ER stress, leading to an increase of prion protein in the secretory pathways[11]. The classical hystopathology of AD consists of extracellular plaques of the amyloid-β peptide and intracellular neurofibrillary tangles of hyperphoshorylated aggregates of microtubule-associated protein *tau*. Aβ-degrading proteases involve a number of subcellular compartments, including lysosomes, endosomes and the ER[12]. Recent results show an overproduction of aggregation-prone amyloid β42, which might interest the ER[13]. This is confirmed by the analysis of the brains of AD patients displaying ER dysfunction[14,15], particularly in the ERAD pathways[16]. Synucleinopathies are a group of neurodegenerative disorders, including Parkinson's disease, that are associated with proteins containing α-synuclein molecules. Experimental evidence suggests that the deposition of α-synuclein in intracellular inclusions together with ubiquitin lead to impaired functions in the ubiquitin-dependent proteasome system[17–21]. Finally, Huntington's disease, an inherited autosomal dominant disease, is caused by the expansion of CAG repeats within the *HTT* gene encoding huntintin, which leads to a protein prone to aggregation in the cytoplasm and alterations in the secretory pathways[22].

The traditional theoretical framework to understand protein polymerization involved in these neurological disorders is based on either molecular dynamics simulations, which give an understanding of how individual proteins interact[23,24], or on kinetic rate equations, which yield the growth of the polymers in a mean-field approximation[25–31]. Theoretical results are then usually compared with *in vitro* measurements of protein aggregation by dynamic light scattering or single molecule fluorescence under well defined conditions of temperature and concentration[30,32–34]. The conditions are, however, not physiological because in the cell proteins undergo a well regulated cycle of synthesis by the ribosome and subsequent degradation through secretory pathways.

In this paper, we demonstrate that taking correctly into account the physiological conditions is important, because rates of synthesis and secretion control a non-equilibrium phase transition for the formation of protein aggregates in the ER. This can be seen in two complementary models. The first is a set of Monte Carlo simulations in a three dimensional (3D) confined space representing the ER, with synthesis and degradation explicitly accounted for. The second is a mean-field model of the same scenario, which offers analytical estimates of long-time protein aggregation. The results from these models illustrate that small variations in control parameters, which could be induced by a variety of biological factors, can lead to large variations in outcomes. This opens a new common perspective on possible diagnostics for these pathologies.

## Results

**Protein polymerization *in vitro*.** We consider a 3D model of linear polymer aggregation in a confined space simulating an intracellular organelle like the ER, a set of channel-like structures surrounding the cell nucleus (see Fig. 1 for an illustration of the model and the Methods section for a detailed description). To validate our model in a well-studied scenario, we first consider the kinetics of a typical *in vitro* experiment and perform 3D numerical simulations of linear polymer diffusion and aggregation at constant concentration and temperature (with $k_{out} = k_{in} = 0$ and periodic boundary conditions as discussed above). While the model is general for linear polymers, we focus here on neuroserpin, which has been shown to undergo processes of activation and latentization[33,34].

We start the simulations from a fixed number $N_m$ of inactive monomers, randomly distributed in space, and then study the effect of different concentrations, measured by the dimensionless density $\rho \equiv N_m L_0^3 / L^3$. We quantify the aggregation kinetics by measuring the weighted polymer mass $m_w = \langle i^2 \rangle / \langle i \rangle$, where $i$ is the number of monomers in each polymer and the average is taken over different realizations of the process. This observable is accessible experimentally through dynamic light scattering[34]. Figure 2a shows that for $k_f = 0$ the weighted mass kinetics displays a crossover between a short-time regime growing as $t^2$, owing to activation and dimerization (Supplementary Fig. 1), and a long-time regime growing as $t^\beta$ with $\beta \simeq 0.5$, owing to polymer–polymer aggregation. The results are in agreement with experiments[34]. We fit the curves obtained at different densities to a crossover function (see Supplementary Information) depending on a crossover time $\tau$ scaling as $\rho^{-\gamma}$ (see the inset of Fig. 2b). The best fit yields $\gamma = 0.149 \pm 0.002$ and allows rescaling of all the curves onto a single master curve. This result confirms that the concentration only sets the timescale of the kinetics. In Fig. 2c, we study the effect of polymer fragmentation showing that if $k_f > 0$ the polymerization process is blocked after a characteristic time that depends on $k_f$. The role of latentization is illustrated in Fig. 2d, which shows that the long-time growth is not affected if $k_L > 0$. Latentization induces a plateau in the crossover region, a feature that is also observed in experiments (see the inset of Fig. 2d and ref. 34). Mean-field theory can also be used to study polymerization kinetics *in vitro*, and yields qualitative agreement with the 3D model, albeit with quantitative differences, as shown in Fig. 3. However, the two models agree on essential features.

**Polymerization *in vivo* displays a non-equilibrium phase transition.** To describe *in vivo* conditions, we introduce to our models non-zero rates for protein synthesis ($k_{in}$) and degradation ($k_{out}$). These effects are in competition: protein synthesis allows greater polymer growth via a flux of monomers that combine into larger polymers; however this growth can be balanced by polymer degradation. To study this quantitatively, we turn first to the mean-field model. Here we characterize polymer aggregation by the mean length $m_p$ of polymers of size $i \geq 2$, which is given by the ratio of the mass and the number of polymers, $\sum_{i \geq 2} i n_i / \sum_{i \geq 2} n_i$. As discussed above, when protein synthesis and degradation are turned off, as in experiments performed *in vitro*, the mean-field model predicts a finite steady state mean polymer size[26]. However, if protein synthesis is turned on, with rate $k_{in}$, then if $k_{out}$ is small, the mean polymer size increases as $Ct^\beta$, with $\beta = 0.5$. On the other hand, if $k_{out}$ is greater than a critical value $k_{out}^c$, which depends on the rates of aggregation and fragmentation, a non-growing steady state is again obtained (Fig. 4a and Supplementary Fig. 2 and 3). As shown in Fig. 4b, we find a sharp transition between growing and stationary phases as quantified by the prefactor of the square-root growth of the mean polymer size $C$, which scales to zero as $(k_{out} - k_{out}^c)^\theta$, with $\theta = 1/2$, when the transition is approached.

Figure 5 shows the phase diagram estimated by the mean-field model as a function of the rates $k_{in}$, $k_{out}$, $k_f$ and $k_p$. The transition region is governed by the products $k_{in}k_p$ and $k_{out}k_f$. Intuitively, $k_{in}k_p$ controls the rate of polymer growth, by the introduction of new monomers and their aggregation into larger polymers, while $k_{out}k_f$ controls the rate of polymer shortening by the removal of material and fragmentation of large polymers. When $k_f$ is large relative to $k_{in}k_p$, the dependence on parameters is further reduced to a dependence on the ratio $k_{in}k_p/k_{out}k_f$. Indeed, in the limit $c = 1$, it can be shown that if a steady state exists, the relationship $(k_{in}k_p)/(2k_{out}k_f) + (k_{in}/k_{out})(1/M_0) = 1$ must be satisfied, where $M_0 = \sum_{i \geq 1} n_i$. This relation determines the location of the critical region as discussed in detail in the Supplementary Information.

We have confirmed the existence of a polymerization phase transition controlled by protein production and degradation by numerical simulations in a 3D system representing the ER, thus mimicking more closely physiological conditions in the cell. As shown in Fig. 4c, 3D simulations show growth curves analogous to the one obtained in the mean-field model, namely $m_p \sim Ct^{1/2}$ (Fig. 4a). In Fig. 4d, we report the prefactor $C$ estimated from the fits in Fig. 4c and observe transition curve similar to the one observed in the mean-field curve (Fig. 4b), with a typical broadening expected for finite size systems.

**Polymerization *in vivo* displays critical-point fluctuations.** To confirm that we are in presence of a critical phase transition, we study the temporal fluctuations of the average polymer length in 3D simulations. As shown in Fig. 6a for a single realization of the process, the polymer length undergoes an intermittent dynamics with bursty activity that is reminiscent of the crackling noise observed in materials close to non-equilibrium phase transitions[35]. We characterize the statistical properties of the signal by measuring the distribution of pulse durations $T$ and sizes $s$, defined as the area under a pulse, for different values of $k_{out}$. Figure 6b shows that for $k_{out} > k_{out}^c$, bursts display a power law distribution with a cutoff that diverges at the transition. In the growing phase, the distribution develops a peak at large $T$ typical of system-wide spanning events. We fit all the subcritical distributions simultaneously by the scaling function

$$P(T, k_{out}) = AT^{-\alpha} \exp(-cT(k_{out} - k_{out}^c)^\delta), \quad (1)$$

obtaining $\alpha = 1.94 \pm 0.02$, $\delta = 2.6 \pm 0.2$ and $k_{out}^c = 1.82 \pm 0.02$ as best fit parameters. Using these values we can collapse the distributions onto a universal scaling function as expected for critical phenomena (see Fig. 6c). A similar analysis for the size distribution yields a power law exponent $\kappa = 1.75 \pm 0.01$, cutoff exponent $1/\sigma = 2.5 \pm 0.2$ and $k_{out}^c = 1.89 \pm 0.02$ (see Supplementary Fig. 4). We notice that these value differ significantly from the predictions of mean-field theory, which give $\alpha = 2$, $\kappa = 3/2$, $\delta = 1$ and $1/\sigma = 2$ (ref. 36).

**Polymerization is system-size dependent.** Three dimensional simulations of protein aggregation display qualitative agreement with mean-field theory in conditions that should apply to experiments in vitro and in vivo. Quantitative differences between the two models are mainly owing to the absence of geometrical features in the mean-field model. Investigating the role of spatial confinement is important because large local fluctuations can drive polymerization in small systems. To this end, we use 3D simulations to estimate the size-dependence of some effective mean-field parameters such as the polymerization and degradation rates, as a function of $k_{out}$ and $H$. The effective degradation rate $\bar{k}_{out}$ depends on the sample size $H$ in three dimensions because proteins can exit only through the boundaries, a feature that is absent in the mean-field model (see Fig. 7). The effective polymerization rate $\bar{k}_p$ displays a peak as a function of $k_{out}$, in correspondence with the phase transition (Fig. 8). Furthermore, for smaller system sizes the polymerization rate is larger.

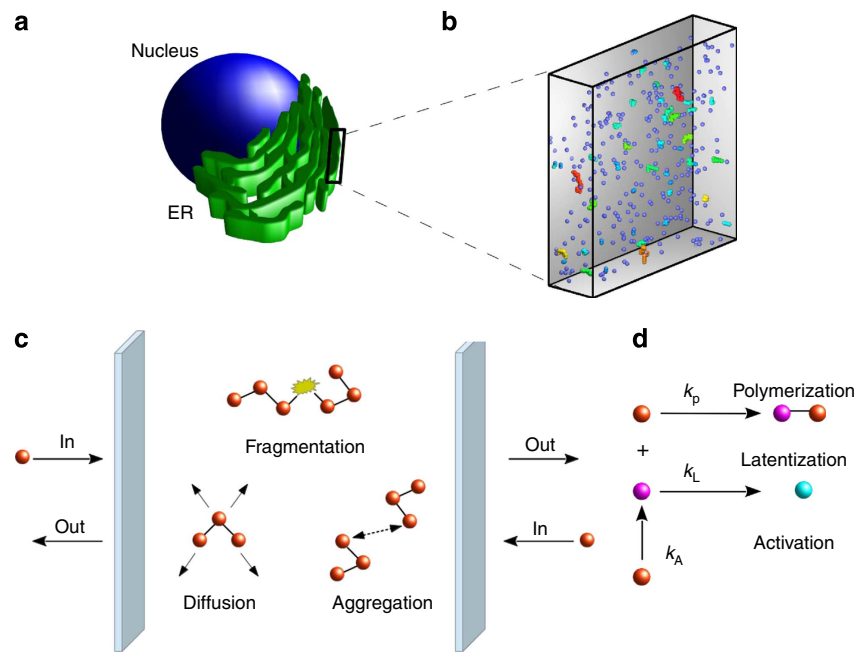

**Figure 1 | Model of protein aggregation in the endoplasmic reticulum.** (**a**) The ER is a set of channel-like structures surrounding the nucleus of the cell. (**b**) We focus on a portion of the ER bounded by two square surfaces placed at distance $H$. Protein monomers are assigned to the nodes of a 3D square lattice. We report a typical configuration obtained from the simulations where polymers are coloured differently according to their length. (**c**) Inside the ER, proteins can diffuse, aggregate and fragment. They can also enter and exit from the ER channel with given rates. (**d**) We also consider the possibility that proteins can aggregate only after becoming active and to transform into a latent state where aggregation is prevented.

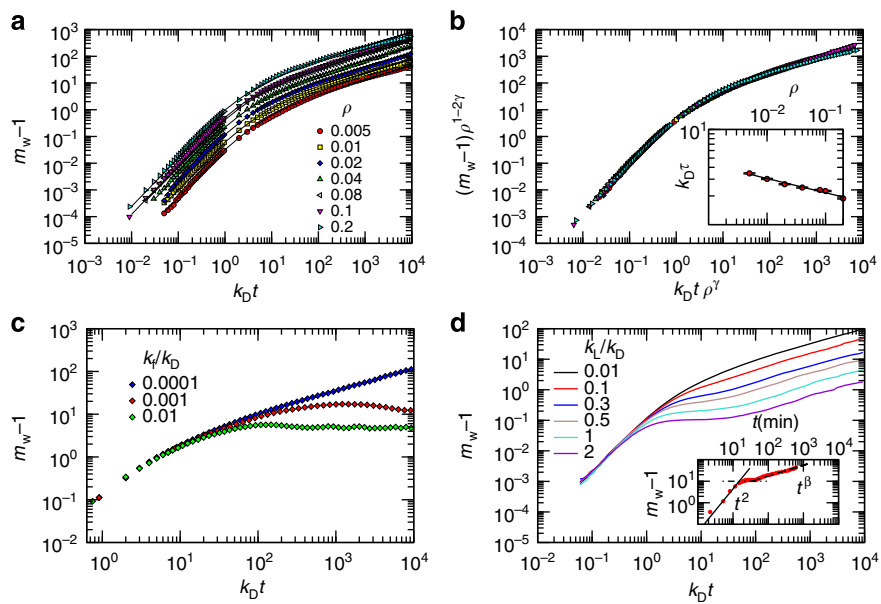

**Figure 2 | Polymerization kinetics in vitro is controlled by protein concentration.** When proteins cannot enter or exit the system, the concentration is constant and controls the kinetics. (**a**) We report the average weighted mass as a function of time for different concentrations for $k_f = 0$ and $k_L = 0$. The curves show a crossover between two power laws and can be fitted as discussed in the main text. (**b**) All the curves from (**a**) can be collapsed into a single master curve when variables are properly rescaled by the concentration. This implies that the crossover timescales as a power law of the concentration as shown in the inset. (**c**) If we use a non-vanishing rate of polymer fragmentation ($k_f > 0$), the growth is limited. (**d**) Latentization ($k_L > 0$) leads to slowing down of the growth, which at long times resumes as in the case $k_L = 0$. This behaviour has been experimentally observed in vitro for neuroserpin, as shown in the inset (data from ref. 34). Curves are obtained by averaging over 10 realizations obtained from statistically identical initial conditions.

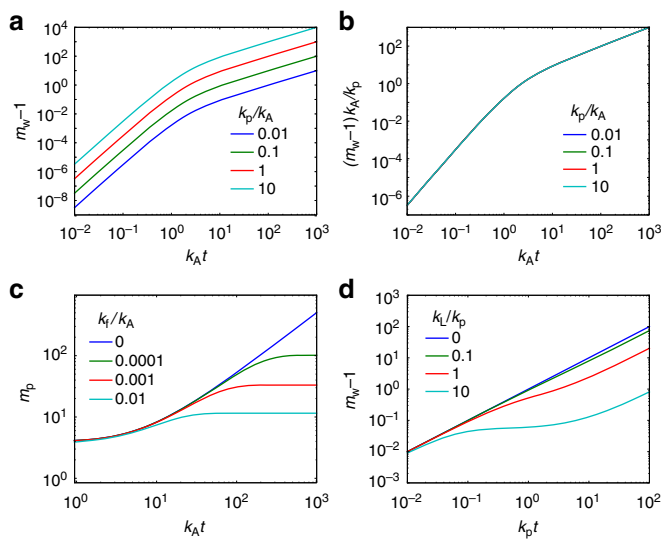

**Figure 3 | Qualitative agreement between mean-field and 3D models.** (**a**) In a model with activation ($k_A = 1$) and no polymer fragmentation or latentization ($k_f = k_L = 0$), the weighted mean polymer size $m_w = M_2/M_1$ exhibits a transition from growth as $t^3$ to linear scaling in $t$. Here we have plotted against time rescaled by activation rate, $k_A t$. (**b**) For fixed $k_A$, time series for different $k_p$ values can be collapsed by rescaling $m_w - 1$ by $k_p$. (**c**) Allowing polymer fragmentation limits the growth of polymers. Here the activation rate is $k_A = 1$, the polymerization rate $k_p = 1$, and there is no latentization ($k_L = 0$), and the growth in unweighted mean size of polymers of length $i \geq 2$, $m_p$ has been plotted. (**d**) Latentization slows polymer growth, as seen here for the weighted mean polymer size $m_w$. Activation is not incorporated here, and no fragmentation occurs ($k_f = 0$), and $k_p = 1$. We plot against time rescaled by $k_p$. When fragmentation is allowed, the system always attains a steady state in which all mass consists of latent monomers. Analogous plots for the 3D model are shown in Fig. 2.

**Comparison with experimental data on amyloid-β clearance from the brain.** Although quantitative experimental data for protein aggregation in the ER are not yet available, the kinetic mechanism we study is general and can be applied also to other situations such as Aβ aggregation and clearance in the brain[12], which has been subject to extensive experimental study[37,38]. In particular, we use our mean-field model to simulate an amino acid labelling experiment like those reported by Bateman et al.[37]. In those experiments, labelled leucine, an amino acid composing Aβ, was injected into the bloodstream of healthy subjects[37]. After a 9 h infusion, the concentration of labelled Aβ in cerebral spinal fluid was monitored for a further 27 h.

To understand experimental results, we simulate the scenario using our mean-field model. We first obtain a steady state by simulating protein aggregation in the steady phase for 50 years. We then follow the labelling experiment of ref. 37 so that over 36 h a fraction of injected monomers are labelled. Labelled monomers evolve following the same dynamics as unlabelled monomers, with full mixing. We calculate the labelled/unlabelled mass fraction in the cerebral spinal fluid, as reported in Fig. 8a.

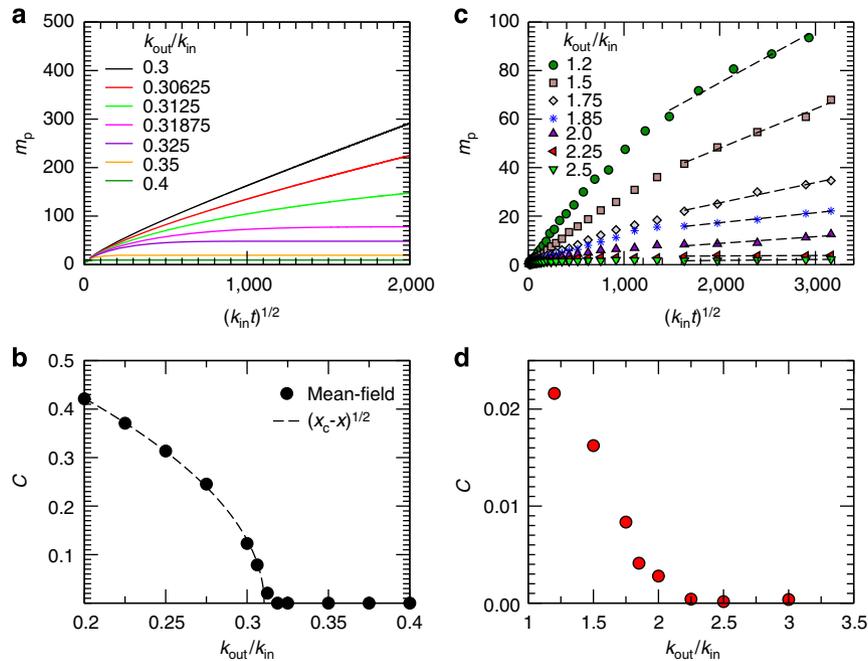

**Figure 4 | Polymerization kinetics *in vivo* displays a non-equilibrium phase transition.** In the ER proteins, enter and exit the system and concentration is not constant. (**a**) Mean-field calculations of the average polymer length for different rate of polymer exit $k_{out}$ grows diffusively as $m_p \simeq C\sqrt{t}$. Above a critical value $k_{out}^c$, which depends on the other rate parameters, growth eventually stops, while it persists for $k_{out} < k_{out}^c$. (**b**) The prefactor $C$ goes to zero at the transition according to a power law with exponent $\theta = 1/2$. (**c**) The average polymer length obtained from 3D numerical simulations grows similar to mean-field theory with a transition at $k_{out}^c \simeq 1.85$. Curves are obtained by averaging over 200–1000 realizations, depending on the proximity to the critical point. (**d**) The prefactor $C$ in three dimensions goes also to zero at the transition but the curve is not sharp, probably because of finite size effects.

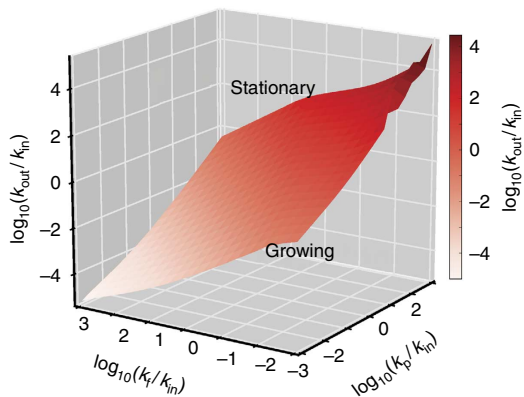

**Figure 5 | Mean-field phase diagram.** Phase diagram of the mean-field model. The manifold in parameter space separates those parameters for which a steady state is obtained (the stationary phase) and those that lead to continual growth of polymers (the growing phase). The model used here has a polymer degradation rate that scales with polymer size $i$ as $i^{-3}$, up to a cutoff $c = 5$, but similar phase diagrams are obtained for different size dependences and cutoffs and in much of the phase space the location of the transition is predicted well by a model with $c = 1$, as discussed in the Supplementary Information.

From this, we determine the fractional clearance rate (FCR) of labelled proteins, as described in the Supplementary Methods (see Supplementary Fig. 5).

Measured FCRs depend on many parameters (see Supplementary Table 1) but here we focus on the role of protein degradation, controlled by $k_{out}$. A single FCR measurement does not allow a determination of all parameters, but our results indicate that a series of labelling experiments can reveal changes in degradation rates over time and the approach to the physiological–pathological phase transition. We take the steady-state system and decrease $k_{out}$ slightly, and allow it to evolve a further 3 years to attain a new steady state, after which we perform another labelling test. As reported in Fig. 9b, a reduced $k_{out}$ yields a reduction in FCR, as well as the maximum labelled/unlabelled ratio. Further decreases in $k_{out}$ cause measured FCR to continue to decrease, but as the critical point is approached the lifetime of labelled proteins diverges and near the critical point no clearance is observed in experimental timescales. Further experiments by the same group indeed reveal that FCR decreases for AD patients with mild symptoms[38], in agreement with the predictions of our model.

### Discussion

Conformational diseases, such as Alzheimer's and Parkinson's diseases, spongiform encephalopathies and serpinopathies, are all associated to aberrant protein aggregation in which the secretory activity of intracellular organelles plays a critical role. In this paper, we have shown by 3D numerical simulations and mean-field calculations that protein aggregation undergoes a non-equilibrium phase transition controlled by the rates of protein synthesis and degradation. Our theoretical analysis compares well with experimental results both *in vitro* and *in vivo*. We can describe with great accuracy the time-dependent polymerization of neuroserpin *in vitro*[34] and experimental measurements of Aβ clearance from the central nervous system in healthy subjects (or with mild AD symptoms)[37,38]. We predict that Aβ clearance measured experimentally in AD patients should decrease as the disease progresses and vanish as the non-equilibrium phase transition is approached. At this point, protein polymerization would not stop. We believe that combining quantitative clinical information with realistic theoretical modelling could be useful to improve diagnostic tools for these pathologies.

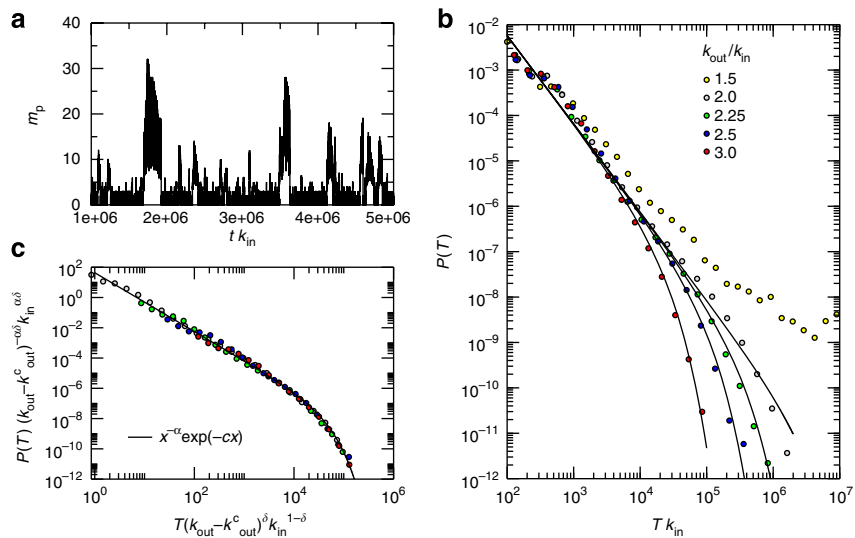

**Figure 6 | Polymer sizes display critical fluctuations close to the phase transition.** (**a**) The average polymer length in a single simulation displays fluctuations with intermittent bursts that are reminiscent of crackling noise. (**b**) The distribution of burst durations follows a power law distribution up to a characteristic cutoff that increases as the transition is approached. Above the transition, we observe large events that should be limited by system size. (**c**) Different distributions can be collapsed onto a universal scaling function.

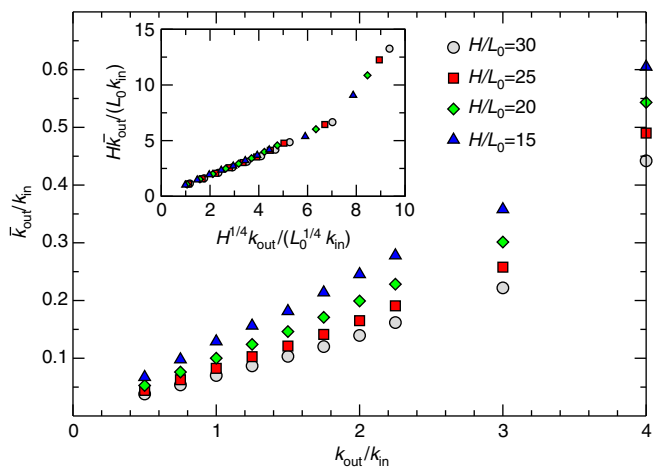

**Figure 7 | Effective degradation rate from numerical simulations.** The effective degradation rate is computed from 3D simulations for different values of $k_{out}$ and $H$. A data collapse is reported in the inset.

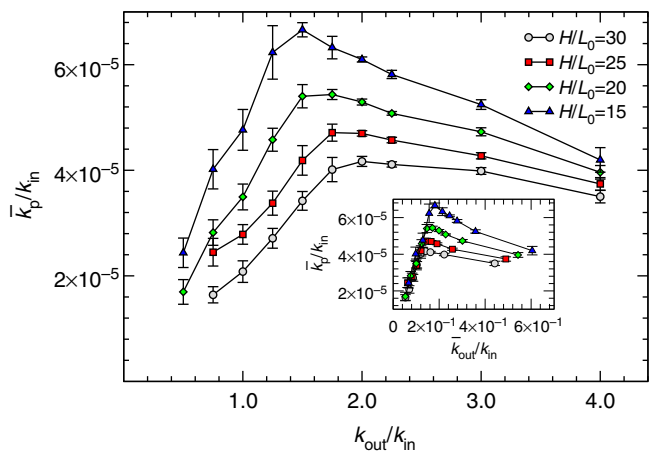

**Figure 8 | Effective polymerization rate from numerical simulations.** The effective polymerization rate is computed from 3D simulations for different values of $k_{out}$ and $H$. The peak corresponds to the phase transition. The same data plotted as a function of $\bar{k}_{out}$ show a collapse only below the phase transition (inset).

In our model, the phase transition is associated with a breakdown of homeostasis in intracellular organelles, controlled by the rates $k_{in}$ and $k_{out}$. This result naturally leads to two fundamental questions: (i) which specific biological processes tune these control parameters and (ii) what are the biological consequences of the transition?

A possible answer to the first question might involve lipid metabolism, which has been recently shown to play a role in all these neurodegenerative diseases[39–42]. In particular, ref. 42 shows that in FENIB the inhibition of HMGCoA reductase, the limiting enzyme of the cholesterol biosynthetic pathway, has a critical role in the clearance of mutant neuroserpin from the ER. Numerous studies have also reported that the modification of cholesterol content can affect amyloid precursor protein processing, which is needed for neuronal activity[40]. Interestingly, a recent paper shows that E693D (Osaka) mutation in amyloid precursor protein promotes intracellular accumulation of Aβ, reducing its excretion[43]. This paper also shows the importance of Aβ trafficking for intracellular cholesterol transport and efflux and that the Osaka mutation potentiates cholesterol-dependent exacerbation of intracellular Aβ toxicity by disturbing amyloid-mediated cholesterol efflux from the cell[43]. These observations can be translated in our model considering that the alteration of lipid metabolism (that is, the level of cholesterol) should lead to a reduction of the parameter $k_{out}$ eventually triggering the phase transition.

As for the second question, our model allows the interpretation of the emergence of conformational diseases in the framework of phase transitions and critical phenomena. This implies that a minimal change in the control parameters can lead to drastic changes in the system when we cross the transition line leading to the disease. The continuous nature of the transition implies, however, that large fluctuations are expected as we approach it.

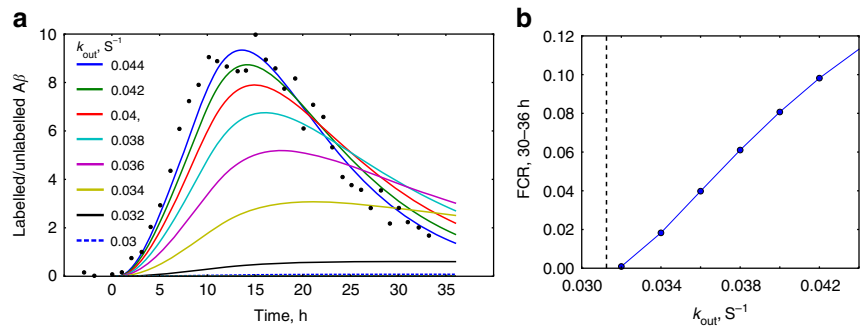

**Figure 9 | Phase transition affects protein clearance from the central nervous system.** Evolution of the measured protein clearance as the physiological–pathological phase transition is approached, simulated in the mean-field model and compared with experimental data. (**a**) Ratio of labelled to unlabelled protein in the cerebrospinal fluid at different values of the control parameter $k_{out}$ (measured in s$^{-1}$). After each measurement, $k_{out}$ is decreased and the system evolves for 3 years to reach a new steady state, as described in the Supplementary Information. Parameter values are chosen so the first labelling measurement agrees with experimental data (shown as black dots), which are taken from Fig. 3c in ref. 37. (**b**) Measured fractional clearance rate (FCR), defined as the slope of the log of the labelled/unlabelled ratio in the period 30–36 h. FCR decreases with $k_{out}$, and near the critical value of $k_{out}$ (here $k_{out} = 0.03125$ s$^{-1}$, indicated by the dashed line), no clearance would be measured before $t = 36$ h.

In perspective this might provide the basis for early detection strategies before the pathology has taken place.

It is important to remark that the phase transition we observe does not occur when the concentration is held constant as usually assumed in experiments *in vitro*. In this case, the response to the variation of control parameters is smooth: small changes of the rates correspond to small changes in the polymerization kinetics. This might be one of the reason behind the difficulty in extrapolating *in vitro* results to the *in vivo* situation. One possibility to overcome this problem would be to use microfluidic devices where the influx and outflux of proteins are externally controlled[44]. Using a similar experimental device, it would be possible to reproduce more faithfully protein aggregation in the cell.

## Methods

**3D model.** In the model, individual proteins are modelled as monomers sitting on a 3D square lattice. Monomers diffuse with rate $k_D$ and attach to neighbouring monomers or polymer endpoints with rate $k_H$. Polymers move collectively by reptation with a length-dependent rate $k_R/i^2$, where $i$ is the number of monomers in the polymer (see ref. 45 p. 89), and locally by end rotations with rate $k_E$ and kink moves with rate $k_K$ (for a review of lattice polymer models see ref. 45). A polymer can attach to another polymer with rate $k_H$ if their endpoints meet, and can fragment by breaking an internal bond with rate $k_f$ (Fig. 1c). Inspired by experimental results on neuroserpin polymerization[34], we allow for polymerization after at least one of the monomers has been activated with rate $k_A$. Active monomers can also become latent with rate $k_L$ and after that they do not aggregate (Fig. 1d).

We consider two types of boundary and initial conditions for the polymerization kinetics. (i) To simulate experiments *in vitro*, we start with a constant number of inactive monomers in a cubic system of size $L = 60L_0$, where $L_0$ is the typical monomer diameter, with periodic boundary conditions in all directions. (ii) To simulate polymerization in the ER, we consider a system of size $(L \times L \times H)$ with $L = 100L_0$ and $H = 25L_0$ with periodic boundary conditions along x and y and closed boundary conditions along z. This choice is justified by the structure of the ER in which a channel of small width (that is, $H$) is bounded by two extended membrane sheets. Monomers enter the system from both closed boundaries with rate $k_{in}$ and monomers and polymers can exit from the same boundaries with rate $k_{out}/i^3$. The fact that the exit rate decreases with the polymer length is suggested by experiments showing that proteasome degradation is slower for larger aggregates[46,47], although no specific measurements exist for the ER. We have checked that different functional dependencies of the exit rate on the number of monomers in the polymer yield similar results.

We perform numerical simulations using Gillespie Monte carlo algorithm[48]. In case (i) we measure time in units of $1/k_D$, setting $k_H = k_E = k_R = k_K = k_A = k_D$ and varying $\rho$, $k_L$ and $k_f$. In case (ii) we measure time in units of $1/k_{in}$ and set $k_H = k_E = k_R = k_K = k_A = k_{in}$, $k_L = 0$, $k_D = 10^2$, $k_f = 10^{-3}k_{in}$ and vary $k_{out}$.

**Mean-field model.** In the mean-field model, the evolution of the populations $n_i$ of polymers of size $i$ is described by a set of coupled nonlinear differential equations, which are based on the assumption that polymer aggregation and fragmentation have no spatial dependence. We have used the model introduced by Blatz and Tobolsky[26] in which polymers aggregate with rate controlled by parameter $k_p$ and fragment with rate controlled by $k_f$. We add two ingredients to this model. The first is production of monomers, which enters as a term $+k_{in}$ added to $\dot{n}_i$, the rate of change of monomer population. The second is polymer degradation, which occurs at a rate that decreases with polymer size. This is described by a set of terms $-k_{out}n_i i^{-3}$. The population of polymers of size $i$, $i \geq 1$, is given by the differential equation

$$\dot{n}_i = \frac{1}{2}k_p \sum_{j=1}^{i-1} n_j n_{i-j} - k_p n_i \sum_{j=1}^{\infty} n_j - k_f n_i(i-1) \\ + 2k_f \sum_{j=i+1}^{\infty} n_j - k_{out}f(i)n_i + k_{in}\delta_{i,1}, \quad (2)$$

where $f(i) = i^{-3}$ for $i \leq c$ and 0 otherwise. Rate equations for the moments of the size distribution and equations for $\dot{n}_i$, $i \leq c$ are solved numerically using standard techniques. The cutoff $c$ is introduced only for numerical convenience, because otherwise it would be impossible to solve the equations in closed form, but we have observed rapid convergence to a $c$ independent solution when $f(i)$ decreases faster than $i^{-2}$. The limit $c = 1$ may have biological significance since it implies that only monomers can exit the system, which implies that polymers have to be fragmented before being degraded. A full discussion of these equations is given in the Supplementary Methods.

### Acknowledgements

We wish to thank M.J. Alava, S. Caracciolo, L. Del Giacco, L. Pollack, J.P. Sethna for useful discussions and suggestions. Z.B., G.C. and S.Z. are supported by ERC Advanced Grant SIZEFFECTS. C.A.M.L.P. and S.Z. wish to thank the visiting Professor Programme of Aalto University where part of this work was completed.

### Author contributions

C.A.M.L.P. and S.Z. designed the research, Z.B. solved the mean-field model, G.C. performed numerical simulations, Z.B., G.C. and S.Z. analyzed data, C.A.M.L.P. and S.Z. wrote the paper with input from ZB and GC. ZB wrote the Supplementary Information. G.C. and Z.B. contributed equally to this work.

### Additional information

**Supplementary Information** accompanies this paper at http://www.nature.com/naturecommunications

**Competing financial interests:** The authors declare no competing financial interests.

**Reprints and permission** information is available online at http://npg.nature.com/reprintsandpermissions/

**How to cite this article:** Budrikis, Z. *et al.* Protein accumulation in the endoplasmic reticulum as a non-equilibrium phase transition. *Nat. Commun.* 5:3620 doi: 10.1038/ncomms4620 (2014).

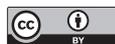